\documentclass[11pt]{article}

\usepackage{amsmath, amssymb, amsthm}
\usepackage{fullpage}
\usepackage{mathrsfs, bbm, natbib}
\usepackage{authblk}
\usepackage{mathtools}
\usepackage{amsmath}
\usepackage{algorithm}
\usepackage{graphicx}
\usepackage[hypcap=true]{caption}
\usepackage{hyperref}
\usepackage{color,soul}
\usepackage{setspace}
\usepackage[utf8]{inputenc}
\usepackage{siunitx} % for units such as km, m^2 a.s.o.

\providecommand{\keywords}[1]{\textit{\textbf{Key words}}: #1}

\DeclareMathOperator{\AUC}{AUC}
\DeclareMathOperator{\cov}{Cov}
\DeclareMathOperator{\EX}{\mathbb{E}}% 
\DeclareMathOperator{\poisson}{Po}

\DeclareMathOperator{\unif}{Uniform}

\newcommand{\one}{\mathbbm{1}}
\newcommand{\mcw}{\mathcal{W}}

\begin{document}
%% \bibliographystyle{elsarticle-num}
%% Usually 'natbib' is better and a 'newer version' of this one...
\bibliographystyle{chicago}

\title{Discrete versus continuous domain models for disease mapping} % : a simulation study

\author[1]{Garyfallos Konstantinoudis} \author[2]{Dominic Schuhmacher}
\author[3]{H{\aa}vard Rue} \author[1]{Ben
    Spycher\footnote{Corresponding author; e-mail:
        ben.spycher@ispm.unibe.ch}} \affil[1]{Institute of Social and
    Preventive Medicine (ISPM), University of Bern, Bern, Switzerland}
\affil[2]{Institute for Mathematical Stochastics, University of
    Goettingen, Germany} \affil[3]{CEMSE Division, King Abdullah
    University of Science and Technology, Saudi Arabia}

\date{} \setcounter{Maxaffil}{0}
\renewcommand\Affilfont{\itshape\small}
\maketitle

\normalsize

\begin{abstract}
    \vspace*{-.5em} The main goal of disease mapping is to estimate disease risk and identify high-risk areas. Such analyses are hampered by the limited geographical resolution of the available data. Typically the available data are counts per spatial unit and the common approach is the Besag--York--Molli{\'e} (BYM) model. When precise geocodes are available, it is more natural to use Log-Gaussian Cox processes (LGCPs). In a simulation study mimicking childhood leukaemia incidence using actual residential locations of all children in the canton of Zürich, Switzerland, we compare the ability of these models to recover risk surfaces and identify high-risk areas. We then apply both approaches to actual data on childhood leukaemia incidence in the canton of Zürich during 1985-2015. We found that LGCPs outperform BYM models in almost all scenarios considered. Our findings suggest that there are important gains to be made from the use of LGCPs in spatial epidemiology.
\end{abstract}
        

\section{Introduction}
Disease mapping, i.e.\ calculating and visualising disease risk across
space, is an important exploratory tool in epidemiology. The
information obtained can provide new clues about the aetiology of a
disease, identify areas of high risk or hotspots, and support
monitoring prevention efforts. Data used for disease mapping
usually consist of disease counts in smaller area units, typically
administrative units such as counties, covering a larger area of
interest. Mapping directly area-level incidence can be misleading, often yielding extreme estimates when the denominator (population at risk) is
small \citep{Wakefield2007}. This problem is usually confronted by exploiting spatial
autocorrelation and borrowing information from neighbouring areas. In
the Bayesian framework a popular class of models are those proposed by
\citet{Besag1991}, often referred to as Besag--York--Molli{\'e} (BYM) models,
which assume global and local smoothing through
conditional autoregressive priors; see \citet{art645} for a
recent treatment. Less frequently, exact geocodes are available,
allowing modelling a disease as a point process over the continuous
spatial domain. An attractive model class of choice in this situation
are the Log-Gaussian Cox processes (LGCPs), among other things because of the tractability of their first and second moments \citep{Moller1998}. Nowadays we have
the computational tools to fit LGCPs in reasonable time but the
additional benefits over the widely used BYM model are not well
understood.

Disease mapping based on areal data is commonly done using the
BYM model, see \citet{Halonen2016, Riesen2018} for examples. The BYM model is an extension of the ICAR (Intrinsic Conditional Autoregressive) model, obtained by adding a spatially unstructured random effect to the already given spatially structured random effect. The latter is a realisation of a Gaussian Markov random field (GMRF) with zero mean and a sparse precision matrix capturing strong spatial dependence~\citep{book80}. The unstructured random effect may be seen as a collection of independent random intercepts for the various areal units. This specification leads to a piecewise constant risk surface which depends on the spatial unit selected and assumes uniform risk
across this spatial unit. Advances in Bayesian inference using integrated nested Laplace
approximations (INLA) have made this method widely accessible and
investigators can get quickly posterior estimates \citep{Rue2009,Illian2012,Rue2017,art643}. The combination of
easy accessible data and freely available code with a toolbox
\citep{art527} have contributed
significantly to the popularity of the BYM model~\citep{art517}.

When precise geocodes are available, it is more natural to study the
point pattern using spatial point process models, see \citet{Diggle2005a, Diggle2013} and \citet{Giorgi2016} for examples in disease mapping. LGCPs model
locations of cases (geocodes) as an inhomogeneous Poisson process
conditional on a latent field, which is a realisation of a Gaussian
random field (GRF) \citep{Moller1998,Illian2012,art589}. In order to do computations, the GRF is often discretized to a regular grid. The covariance matrix of
the discretized field has an intuitive interpretation, but is typically a dense matrix, leading to high computation costs (big $n$ problem \citep{Lasinio2013}). Computational techniques can be exploited which make this procedure tractable, but when combined with Monte
Carlo algorithms the computational burden remains large. Advances
include more efficient inferential tools that use better proposal mechanisms \citep{Girolami2011}, INLA \citep{Rue2009} or different approximations of the
covariance matrix \citep{Heaton2017}. Recently, \citet{Lindgren2011} proposed a finite element based approximation to the stochastic weak solutions of the stochastic partial differential equations (SPDE) that describe certain GRFs with Mat{\'e}rn covariance function \citep{Whittle1954}. This approach
allows to specify an arbitrary triangulation of space and yields a GMRF
representation of the (approximate) solution indexed by the vertices; see \citet{art643} for a recent review. This is more appealing than the dense LGCP approach described above, since the Markov property allows to do computations based on a sparse precision matrix, while keeping the continuous GRF model without an artificial specification of a regular grid; see \citet{Pereira2017} for an example.

The continuous nature of LGCPs leads to several preferable theoretical
characteristics compared to the BYM models. First LGCPs are resolution
invariant, i.e.\ they bypass all the problems arising when dealing with
arbitrary boundaries; for example, the modifiable areal unit problem,
where the results are highly dependent on the areal unit selected
\citep{openshaw1984}. Inference for BYM is also complicated by numerous
irregular changes in the regions on which health data is reported \citep{Li2012a}.
In addition, BYM assumes constant risk within the spatial units, but in most situations the unknown spatial covariates associated with the disease of interest are
expected to be continuous, making this starting point a strong
assumption. Furthermore, if the areas of higher risk are smaller than
the areal unit selected, the BYM model is not expected to be as
sensitive and specific as a continuously indexed model. Lastly
covariates are often available at different spatial scales. LGCPs
allow using all the data sources available, retaining high-resolution
and overcoming problems such as spatial misalignment and ecological
bias \citep{Gotway2002}. These preferable theoretical characteristics coupled with the fact that aggregating point data into regional counts results in an information loss suggest that LGCPs should outperform the BYM model. But is this true in practice and how can we quantify any such improvement? 

There are a few published studies that compared these methods.
A study examining lupus incidence in Toronto, simulated 40 Gaussian
random fields using a Mat{\'e}rn correlation function with roughness and
variance parameters fixed, varying the range parameter \citep{Li2012}.
They compared the models’ ability to calculate the risk and identify
areas of higher risk and concluded that LGCPs outperform BYM in all
instances. Using similar simulation procedure and metrics,
\citet{Li2012a} extended the LGCP model, assuming that exact case
locations are unknown and information is only available at larger area
units (census tracks in their example), and compared this version with
the BYM model. They reported that their LGCP version
outperforms the BYM model, however when case locations are available, it
is preferable to use LGCP on the exact points rather than LGCP on
aggregated data. It is not surprising though that in both studies
LGCPs performed best, given that the processes used to generate and fit the data 
(Mat{\'e}rn with roughness parameter 2) were the same. An Australian study
using 6 scenarios consistent with a previous study \citep{Illian2012}
assessed the performance of, among other models, the BYM and LGCP with a
Mat{\'e}rn correlation function on different spatial scales
\citep{Kang2013} by assessing the deviance information criterion (DIC)
and the logarithmic score. They concluded that the models' prediction performance was 
scenario dependent and suggested that the analysis should be performed using different 
spatial scales and thus smoothness priors. However, they did not examine their ability to
identify areas of higher risk. All three studies were based on a small
number of datasets and none incorporated the continuous (triangulation-based)
specification of the precision matrix by \citet{Lindgren2011}.

Today, more than ever before, geo-referenced data are available at high spatial resolution. Nevertheless, due to confidentiality concerns, such data are often aggregated in some spatial unit. This aggregation leads automatically to the use of a BYM-type model. The goal of our investigation is to compare the pairs BYM with areal data and LGCP with point data to examine to what extent the availability of individual data and use of an LGCP model has practical benefits. In addition, we wanted to assess the performance of the pair LGCP and SPDE as a toolbox for disease mapping compared to the most popular disease mapping method. We investigated the performance of BYM and LGCP when the interest lies in quantifying risk across space (mapping) and identifying areas of increased risk. For this we perform an extensive simulation study based on a real spatial population. Our findings are then used to interpret the BYM and LGCP model fits for the childhood leukaemia incidence during 1985-2015 in the canton of Z{\"u}rich. The remainder of the paper is laid out as
follows. Section \ref{sec:Methods} describes the methods used in this article, how data
was simulated and what metrics are used to assess the performance. In Section 3
we present and discuss the results of the simulation study, whereas in Section 4 the
models are applied to the childhood leukaemia incidence in the canton of
Zürich. Section 5 gives a general discussion and areas for future
work and section 6 ends with the conclusion.

\section{Methods}
\label{sec:Methods}
	\subsection{Models}

Let $\mathcal{W}$ be an observation window subdivided in spatial units
$A_1, \dots, A_N$ and denote by $Y_i$ be the disease count in the
$i$-th unit. Suppose that $Y_i\sim\poisson(\lambda_iP_i)$, where $P_i$
is the population in the $i$-th spatial unit and $\lambda_i$
the corresponding risk. The BYM model specification assumes:
\begin{equation} \label{CARmodel}
\begin{split}
\log(\lambda_i) &= \beta_0 + u_i + v_i \\
u_{i}\,|\,\pmb{u}_{-i} &\sim \mathcal{N}\biggl(
\frac{\sum_{j=1}^Nw_{ij}u_j}{\sum_{j = 1}^{N}w_{ij}},
\frac{1}{\tau_1\sum_{j=1}^{N}w_{ij}}\biggr) \\
v_i &\sim \mathcal{N}(0, \tau_2^{-1})
\end{split}
\end{equation}
where $\beta_0$ is a constant, $u_i$ is a spatially structured random
effect (ICAR component; $\pmb{u}_{-i}$ denotes $(u_j)_{j \neq i}$), and $v_i$ is a spatially unstructured random effect (independent random intercepts for different~$i$). The $w_{ij}$ represent weights taking the value 1 when spatial units
$i$ and $j$ are first order neighbours and 0 otherwise, and
$\tau_1$ and $\tau_2$ denote random precision parameters. Specifying
appropriate priors for the precision parameters completes the
Bayesian representation of the above model. Following the
parametrisation by \citet{Simpson2017} and \citet{art585} the above
equation is rewritten as:
\begin{equation} \label{CAR_2} \log(\lambda_i) = \beta_0 +
\frac{1}{\sqrt{\tau}}\bigg(\sqrt{1-\phi} v_i + 
\sqrt{\phi}u_i^*\bigg)
\end{equation}
where $v_i \sim \mathcal{N}(0, 1)$, $u_i^*$ is a standardised spatial component that has characteristic
marginal variance equal to 1 \citep{art521}, $\phi\in[0,1]$ is a mixing
parameter and $\tau$ controls the marginal precision. Using the
representation given in \eqref{CARmodel} leads to an independent
assignment of priors on the precision parameters, which may lead to identifiability issues for the case where no spatial dependence is found \citep{Simpson2017, MacNab2011}. In \ref{CAR_2} the hyperparameters $\phi$ and $\tau$ are orthogonal in interpretation, which allows us to specify priors
independently.

Turning now to the continuous domain, let $Y$ be a an
inhomogeneous Poisson point process on $\mathcal{W}$ with mean expected
number of points in any set $A\subset\mathcal{W}$ equal to
$\int_{A}p(s)\lambda(s) \, ds$, where $p(s)$ is the population density and
$\lambda(s)$ is the risk at location $s$ \citep{art583}. In an LGCP model
we assume that the log-risk $\log \lambda(s)$ (and hence the log-intensity
of $Y$) is the realisation of a Gaussian random field
$Z = (Z_s)_{s \in \mathcal{W}}$. Assuming stationarity and isotropy yields the model specification:
%
% This equation is currently not referenced; but it will make our life much
% easier if we can refer to it (in communications, later papers and so on)
\begin{equation} \label{LGCPmodel}
\begin{split}
\log \lambda(s) &= \beta_0 + Z(s) \\
\EX[Z(s)] &= 0 \\
\cov[Z(s), Z(s+h)] &= k(h)
\end{split}
\end{equation}
where  $k(\cdot)$ is a symmetric non-negative definite function depending on the marginal variance $\sigma^2$ and a range parameter $\varrho$, beyond which correlations fall below a certain threshold of approximately~$0.1$. The LGCP specification~\eqref{LGCPmodel} allows for the inclusion of covariates via further additive terms in the first equation; the same holds for the BYM specification~\eqref{CARmodel}. Typical choices for $k(\cdot)$ include the exponential, Gaussian, and spherical covariance functions. For this particular approach we used the popular and very flexible class of Mat{\'e}rn covariance functions, which has an additional roughness parameter $\nu$ that is fixed (determined by the investigator). Following \citet{Lindgren2011}, we assume a finite element representation of the Mat{\'e}rn field based on a fairly dense triangulation referred to as mesh (online supplement, Figure~S1):
\begin{equation}\label{finite}
Z(s) \approx \sum_{i = 1}^M \psi_i(s) Z_i,
\end{equation}
where $M$ denotes the total number of mesh nodes, $Z_i$ are random weights and $\{\psi_i\}$ is a set of piecewise linear basis functions taking the value $1$
at the $i$-th mesh node, and $0$ at every other node.
\citet{Whittle1954,Whittle1963} showed that the solution $Z(s)$  of the stochastic partial
differential equation
\begin{equation}\label{spde}
\theta (\kappa^2 - \Delta)^{\alpha/2}Z(s) = W(s)
\end{equation}
is a GRF with Mat{\'e}rn covariance function under the reparametrization
\begin{equation*}
\alpha = \nu + d/2, \quad \kappa = \sqrt{8\nu} \, \varrho^{-1}, \ \text{and} \quad 
\theta^2 = \frac{\Gamma(\nu)}{\Gamma(\nu+d/2) (4\pi)^{d/2} \kappa^{2\nu}} \, \sigma^{-2},
\end{equation*}
where $d$ is the dimension of the space. Here $W(s)$ denotes Gaussian white noise and $\Delta=\sum_i \partial^2/\partial s_i^2$ is the Laplacian. For this analysis we use $\nu=1$. Computing an approximate stochastic
weak solution of \eqref{spde} based on the finite element representation~\eqref{finite}
results in a Gaussian vector $\mathbf{Z} = (Z_i)_{1 \leq i \leq M}$ with mean zero
and sparse precision matrix $\mathbf{Q}(\theta,\kappa)$.
Unlike traditional methods for inference in LGCP models, this appoach uses
the precise locations in the point pattern without aggregation and
provides a continuous approximation of the latent field.

\subsection{Data simulation}

To compare the performance of the two models described above, we
conducted a simulation study. In this section, we describe the data
simulation procedure.

The selection of scenarios was motivated by the example of childhood
leukaemia incidence in Switzerland. Childhood leukaemia is a rare
cancer and over the period 1985--2015 we observed $n=334$ childhood
leukaemia cases in the canton of Zürich, which had a total childhood
population ($<16$ years of age) of $P_{\mcw} = 206,\!532$ in 2000. Precise geocodes were
available from the national census in 2000 allowing to simulate case
locations from the true underlying geographic distribution of the
population at risk.

\begin{figure}[t]
	\centering \includegraphics[width=65mm]{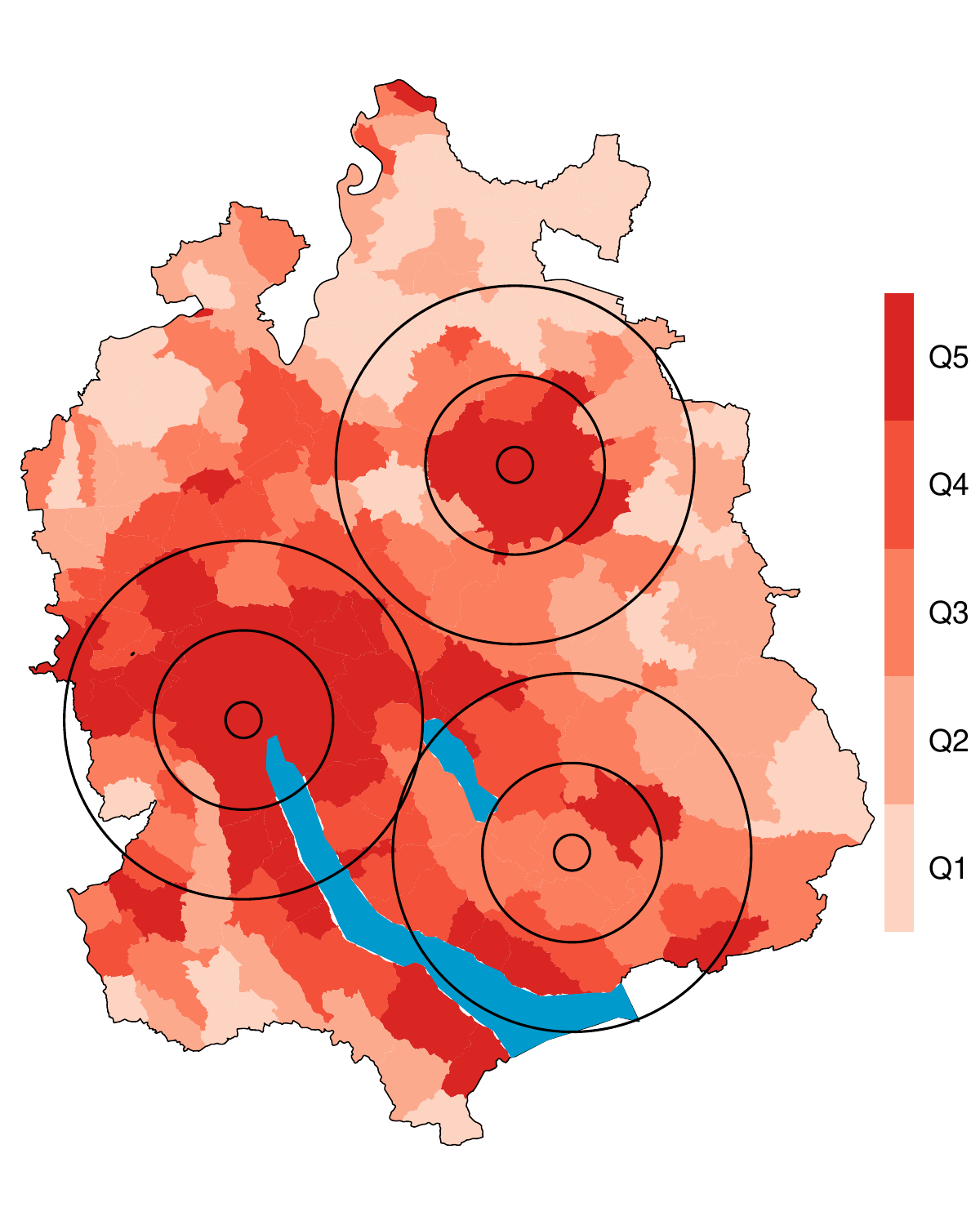}
	\caption{The circular high risk areas considered in the
		simulation study (radii = 1, 5 and \SI{10}{km}). The shading shows the population
		density per municipality in quintiles in the canton
		of Zürich based on data of the 2000 census. The population density here refers to children $<16$. Although we used the precise geocodes in our main analysis, data confidentiality considerations do not allow us to show the childhood population density on a finer geographical scale.}
	\label{fig1}
\end{figure}

We considered scenarios varying in the size of high-risk areas
(radius $r$ of circular high risk areas in \si{km}; $r\in\{1,5,10\}$), the
risk ratio between the low risk area and the high risk
area ($c\in\{2,5\}$), the expected number of cases generated ($kn$, where
$k\in\{1,5,10\}$ with $n=334$ from above) and the shape of the risk surface
(step function or smooth function). All of the resulting 36 scenarios
included 3 high risk areas with centres located in a highly urban area
(Zürich; Figure~\ref{fig1}, circles on the left), a semi-urban area
(Winterthur; Figure~\ref{fig1}, top-right circles) and a highly rural area
(Gossau; Figure~\ref{fig1}, bottom-right circles).
We also included 3 scenarios with a flat risk surface for $k\in\{1,5,10\}$.
For each of the resulting 39 scenarios, we generated 300 datasets.

We selected a circular shape for the high risk areas because of its
simplicity (defined only by centre and radius), rotational
invariance (thus avoiding arbitrary choices of angular orientation),
and because it can be regarded as a generic model of environmental
contamination from a point source. Furthermore it is unlikely to
favour any of the models by unintentional alignment with the
subdivisions of space used in model fitting, i.e.\ municipalities for
BYM or a Voronoi tesselation or regular grid for LGCP models.

In the scenarios for which the true risk surface is a step function set
\begin{equation*} 
\lambda_{\text{step}}(s) = \lambda_0 \bigl(1 + \alpha \max_l \one\{\|s-x_l\| \leq r\} \bigr), \quad s \in \mcw,
\end{equation*}
where $\lambda_0$ is the risk outside the circles,
$\alpha = c - 1$ is the proportion of the excess risk inside the circles, 
$x_l$ is the centre of the $l$-th circle, $l=1,2,3$,
and $\one\{ \text{condition} \}$ takes the
value 1 if the condition is satisfied and 0 otherwise. The risk at the location
of residence $s_i$ of the $i$-th child is then given by $\lambda_i = \lambda(s_i)$
for $i=1,\ldots,P_{\mcw}$.

For each value of $c$ and $k$, the baseline risk $\lambda_0$ was selected such that
the overall number of expected cases generated would equal $kn$. To generate case locations, we
sampled a value from $\unif(0,1)$ for each person $i = 1, \ldots, P_{\mcw}$, and declared the person to be a case if the sampled value was smaller than~$\lambda_i$. We thus generated $J=300$
datasets. The full algorithm used to generate the datasets is given in
the online supplement as Algorithm S1.

In the scenarios with a smooth risk surface the excess
risk was modelled using Gaussian functions as follows:
\begin{equation*} 
\lambda_{\text{smooth}}(s) = \lambda_0 + \beta \max_l
\Big\{\exp \Bigl(-\frac{\|s- x_l\|^2}{2\gamma^2}\Bigr)\Big\}, \quad s \in \mcw,
\end{equation*}
where $\lambda_0$ denotes the background risk and $x_l$ are as above. While taking the sum of the three Gaussian components may seem more intuitive, we selected the $\max$, because this way the shape of the high risk areas remains intact (clear circles). 
For each combination of
$c$ and $r$, we selected the new parameters $\beta, \gamma > 0$ such that
a) on average 80\% of the excess cases produced by an isolated Gaussian risk
function over an infinite area occur within a circle of radius
$r$;
b) the expected number of excess cases produced by the risk
surface $\lambda_{\text{smooth}}$ over the canton of Zürich is the same as under
$\lambda_{\text{step}}$, and c) the expected total number of
cases is the same under both risk surfaces.
To sample locations we used the same procedure as described above. For
more information how $\gamma$ and $\beta$ were derived, for the sampling
algorithm and a graphical representation of the risk surfaces under
different scenarios, refer to the online supplement, Section 1,
Algorithm S2 and Figures S2--4.

\subsection{Prior selection and inference}

Both for the BYM and LGCP models and across all datasets in the simulation, we followed the results
from \citet{Simpson2017} to construct penalised
complexity priors. These priors are invariant to
parametrisations, have a natural connection with Jeffrey's priors, are
parsimonious and have excellent robustness properties \citep{art590,
	Sorbye2017, Simpson2017}. For the BYM model we set a prior for
$\tau$ in \eqref{CAR_2} such that $\Pr(1/\sqrt{\tau}>1)=0.01$ indicating that the log-risk in a fixed area is unlikely to have variance more than $1$. For the mixing parameter $\phi$ we assigned
$\Pr(\phi \leq 0.5) = 0.5$ implying that the median of the mixing parameter is 0.5 (i.e.\ equal contribution of the overdispersion component and the ICAR component to the latent field).
For the LGCP model we followed a similar approach for the
marginal standard deviation, setting again $\Pr(\sigma>1)=0.01$, whereas for the
range parameter we set $\Pr(\varrho<30000)=0.5$ corresponding to a weakly
informative prior using the fact that \SI{30000}{m} is roughly half of the diameter of the domain.
Inference for both models was conducted
using INLA as introduced by \citet{Rue2009}; see \citet{book125} for
book-treatment of the subject.

\subsection{Performance measures}

We used the root mean integrated squared error evaluated on a fine
grid as a metric to assess the ability of a model to estimate the true
risk surface:
\begin{equation}
\label{rmise}
\text{RMISE} =
\bigg(\EX\int_\mathcal{W}b(s)(\hat{R}(s)-R(s))^2ds\bigg)^{1/2}
\approx \bigg(\EX\sum_{g=1}^{G}b_g|D_g|(\hat{R}_g-R_g)^2\bigg)^{1/2},
\end{equation}
where $b(s)$ denotes a weight function, $\hat{R}(s)$ is the fitted value at $s$ (a random variable having the marginal posterior distribution) and $R(s)$ is the true value at $s$. For approximating the integral we use on the right hand side the partition $\{D_1, \ldots, D_G\}$ of the domain $\mcw$ into small pixels and $b_g$, $\hat{R}_g$, $R_g$ are suitably chosen representative values of $b(s)$, $\hat{R}(s)$, $R(s)$ on $D_g$, respectively. More precisely, $\hat{R}_g$ is a value simulated from the marginal posterior distribution at $g \approx s$ and the expectation on the right hand side is the average over all such simulated values.
We considered four versions of this RMISE, varying the weights among 
$b_g=1$ and $b_g=\#(\text{people in }D_g)/|D_g|$  where $|\cdot|$ denotes the  area of $D_g$ and the $R$-values among $\hat{R}_g = \log(\hat{\lambda}_g)$ and $\hat{R}_g = \hat{\lambda}_g$,
where $\lambda_g$ is evaluated at the centroid of grid cells. For the
rest of the paper, RMISE refers to the version with $b_g=1$ and
$\hat{R}_g = \log(\hat{\lambda}_g)$ unless otherwise stated.

As a second measure to assess a model's ability to capture the true
risk, we used the coverage probability. Let $\delta_{jg}$ be an
indicator taking the value one whenever $\lambda_g$ lies inside the
95\% credibility region of $\hat{\lambda}_g$ and zero otherwise for the
$j$-th dataset. We defined the coverage probability of the $g$-th cell
as $p_g = \sum_{j=1}^{300} \delta_{jg}/300$. We also
calculated a coverage proportion of cells correctly covered by the
$j$-th map defined as $p_j = \sum_{g=1}^{G} \delta_{jg}/G$.
For the BYM on municipalities we used the credibility regions of the
municipality, in which the centroid of the grid cell lay.

To assess a model's ability to identify high-risk areas we estimated
the receiver operating characteristic (ROC) curve and determined the
area under this curve (AUC). More
specifically, we defined regions of high risk based on exceedance
probabilities as the set of grid cells satisfying
$\Pr(\hat{\lambda}_g > n/P_{\mcw}) > q$ for some $q \in [0,1)$, where the
probability is taken over the posterior distribution of $\hat{\lambda}_g$.
Denoting the true high risk region, given by $\lambda_g > n/K$,
as $A$ and the region of high risk indicated by the exceedance
probability as $B_\alpha$, we define the area-based sensitivity and
specificity as
\begin{displaymath}
\text{sensitivity}_q = \frac{|A\cap B_q|}{|A|} \quad
\text{and} \quad \text{specificity}_q = \frac{|A^c \cap
	B_q^c|}{|A^c|},
\end{displaymath}
where $|\cdot|$ denotes area and $A^c$ and $B_q^c$ denote the complements of $A$ and $B_q$, respectively; see Figure~S5 in the online supplement for illustration. We evaluate the area-based sensitivity and specificity at $q = 0, 0.05, 0.1, \dots, 0.95$ and calculate $\AUC$ as the area
under the ROC curve defined by plotting sensitivity against $1- $specificity. We also use a population-based version of sensitivity and specificity using the same
formulae as above with $|\cdot|$ denoting population in a given area.
For the rest of the manuscript, $\AUC$ refers to the area-based version
unless otherwise stated.

\section{Results}

	\label{sec:Results}
\begin{table}[!t]
	\caption{\label{tab2} Root mean integrated squared error (RMISE) divided by
		1,000 for $b_g=1$ and $\hat{R}_g = \log(\hat{\lambda}_g)$
		based on \eqref{rmise} in the 36 scenarios with high risk areas.
		BYM stands for the Besag--York--Molli{\'e}
		model, LGCP for the Log-Gaussian Cox process model and $c$ for the
		factor of risk increase within the high risk areas.} 
	\centering
	\begin{tabular}{lllll}
		\hline
		Data generating model & Step function &  & Smooth function &  \\\hline
		Fitted model & BYM & LGCP & BYM & LGCP \\\hline
		\textbf{k=1} &  &  &  &  \\\hline
		Radius = \SI{1}{km} & & & & \\
		c = 2 & 6.76 (4.8, 12.1) & 6.83 (4.5, 12.4) & 6.71 (4.7, 12.2) & 6.76 (4.37, 12.1) \\ 
		c = 5 & 11.8 (7.92, 17.8) & 16.4 (10.5, 21.9) & 12.3 (8.15, 19.4) & 16.2 (10.9, 22.6) \\ 
		Radius = \SI{5}{km} & & & & \\
		c = 2 & 14.8 (12.4, 19.5) & 14.6 (12, 18.9) & 13.6 (10.7, 18.3) & 13.8 (10.9, 18.4) \\ 
		c = 5 & 28.3 (25.4, 33.3) & 26.6 (24.1, 32.3) & 25.1 (22.6, 30.1) & 23.3 (20.3, 28.9) \\ 
		Radius = \SI{10}{km} & & & & \\
		c = 2 & 16.9 (15.1, 19.7) & 14.7 (13.5, 17.9) & 15.4 (13.3, 18.3) & 13.5 (11.6, 17.4) \\ 
		c = 5 & 35.6 (34, 37.6) & 27 (25.4, 29.5) & 27.2 (25.6, 29.4) & 19.8 (18.1, 23.5) \\\hline
		\textbf{k=5} &  &  &  &  \\\hline
		Radius = \SI{1}{km} & & & & \\
		c = 2 & 4.47 (3.17, 6.81) & 6.62 (4.24, 9.88) & 4.48 (3.1, 6.88) & 6.51 (4.27, 9.9) \\ 
		c = 5 & 10.4 (8.77, 12.5) & 14.8 (13.1, 17.1) & 10.8 (8.82, 12.5) & 14.8 (13, 16.8) \\ 
		Radius = \SI{5}{km} & & & & \\
		c = 2 & 11.6 (10.6, 13.1) & 12.2 (10.8, 14.7) & 10.4 (9.32, 12) & 11 (9.33, 14.3) \\ 
		c = 5 & 22.8 (21.4, 24.5) & 21.5 (19.6, 24.6) & 19.2 (18, 20.6) & 16.8 (14.8, 19.9) \\ 
		Radius = \SI{10}{km} & & & & \\
		c = 2 & 14.9 (14.3, 15.8) & 12.1 (11, 14.4) & 12.3 (11.5, 13.4) & 10.1 (8.57, 12.7) \\ 
		c = 5 & 28.4 (27.3, 29.8) & 22.3 (20.8, 24.6) & 21.8 (21, 22.8) & 13.9 (12.1, 17) \\\hline
		\textbf{k=10} &  &  &  &  \\\hline
		Radius = \SI{1}{km} & & & & \\
		c = 2 & 4 (3.01, 5.77) & 7.32 (5.42, 9.68) & 3.99 (2.89, 5.89) & 7.34 (5.43, 9.82) \\ 
		c = 5 & 9.76 (8.65, 11) & 14 (12.8, 15.7) & 9.88 (8.77, 11.1) & 13.9 (12.7, 15.6) \\ 
		Radius = \SI{5}{km} & & & & \\
		c = 2 & 10.4 (9.8, 11.4) & 11.5 (10.2, 13.4) & 9.12 (8.44, 10) & 10.3 (8.66, 12.4) \\ 
		c = 5 & 20.6 (19.6, 21.8) & 19.9 (18.2, 22.8) & 16.9 (16.1, 18.1) & 14.7 (12.9, 17.2) \\ 
		Radius = \SI{10}{km} & & & & \\
		c = 2 & 13.6 (13.1, 14.2) & 11.8 (10.4, 13.9) & 11.1 (10.5, 11.8) & 9.17 (7.75, 11.7) \\ 
		c = 5 & 25 (24.2, 26) & 21 (19.7, 23.3) & 19 (18.2, 19.8) & 11.9 (10.5, 14.8) \\\hline
	\end{tabular}
\end{table}

\begin{table}[!t]
	\caption{\label{tab3} Coverage proportion for the 36 scenarios. BYM stands for the
		Besag--York--Molli{\'e} model, LGCP for the Log-Gaussian Cox
		process model and $c$ for the risk increase within the high risk
		areas. The coverage proportion is defined as the proportion of
		grid cells for which the true risk lies with in the
		credibility region. Given are the median and in parenthesis the
		2.5 and 97.5 percentiles of the mean coverage over the
		simulations.} 
	\centering
	\begin{tabular}{lllll}
		\hline
		Data generating model & Step function &  & Smooth function &  \\\hline
		Fitted model & BYM & LGCP & BYM & LGCP \\\hline
		\textbf{k=1} &  &  &  &  \\\hline
		Radius = \SI{1}{km} & & & & \\
		c=2 & 0.99(0.94,0.99) & 1.00(0.94,1.00) & 0.99(0.95,1.00) & 0.99(0.94,1.00) \\ 
		c=5 & 0.94(0.90,0.99) & 0.99(0.98,1.00) & 0.94(0.91,0.99) & 0.99(0.98,1.00) \\ 
		Radius = \SI{5}{km} & & & & \\
		c=2 & 0.94(0.85,0.97) & 0.97(0.87,1.00) & 0.95(0.93,0.96) & 0.99(0.94,1.00) \\ 
		c=5 & 0.90(0.86,0.94) & 0.95(0.91,0.97) & 0.92(0.90,0.93) & 1.00(0.97,1.00) \\ 
		Radius = \SI{10}{km} & & & & \\
		c=2 & 0.62(0.29,0.99) & 0.90(0.47,0.98) & 0.97(0.54,1.00) & 0.99(0.84,1.00) \\ 
		c=5 & 0.51(0.43,0.91) & 0.82(0.59,0.90) & 0.86(0.62,0.96) & 0.99(0.92,1.00) \\\hline
		\textbf{k=5} &  &  &  &  \\\hline
		Radius = \SI{1}{km} & & & & \\
		c=2 & 0.94(0.91,0.99) & 0.99(0.89,1.00) & 0.95(0.91,0.99) & 0.99(0.88,1.00) \\ 
		c=5 & 0.90(0.88,0.90) & 0.99(0.98,0.99) & 0.90(0.89,0.93) & 0.99(0.98,1.00) \\ 
		Radius = \SI{5}{km} & & & & \\
		c=2 & 0.90(0.85,0.94) & 0.95(0.89,0.97) & 0.92(0.90,0.93) & 0.99(0.93,1.00) \\ 
		c=5 & 0.88(0.84,0.90) & 0.92(0.89,0.94) & 0.87(0.85,0.89) & 1.00(0.96,1.00) \\ 
		Radius = \SI{10}{km} & & & & \\
		c=2 & 0.88(0.52,0.95) & 0.88(0.8,0.94) & 0.93(0.82,0.95) & 1.00(0.94,1.00) \\ 
		c=5 & 0.85(0.76,0.9) & 0.84(0.78,0.88) & 0.85(0.71,0.9) & 0.99(0.96,1.00) \\\hline
		\textbf{k=10} &  &  &  &  \\\hline
		Radius = \SI{1}{km} & & & & \\
		c=2 & 0.94(0.90,0.99) & 0.98(0.85,0.99) & 0.94(0.91,0.99) & 0.99(0.86,1.00) \\ 
		c=5 & 0.90(0.88,0.90) & 0.98(0.98,0.99) & 0.90(0.88,0.90) & 0.99(0.98,0.99) \\ 
		Radius = \SI{5}{km} & & & & \\
		c=2 & 0.89(0.85,0.91) & 0.92(0.85,0.96) & 0.90(0.88,0.91) & 0.97(0.91,1.00) \\ 
		c=5 & 0.87(0.82,0.89) & 0.92(0.88,0.93) & 0.85(0.82,0.87) & 0.99(0.94,1.00) \\ 
		Radius = \SI{10}{km} & & & & \\
		c=2 & 0.88(0.74,0.93) & 0.86(0.79,0.91) & 0.90(0.82,0.93) & 0.98(0.92,1.00) \\ 
		c=5 & 0.86(0.80,0.90) & 0.84(0.79,0.87) & 0.83(0.77,0.86) & 0.99(0.95,1.00) \\\hline
	\end{tabular}
\end{table}

\begin{figure}[h!]
	\centering \includegraphics[width = 150mm]{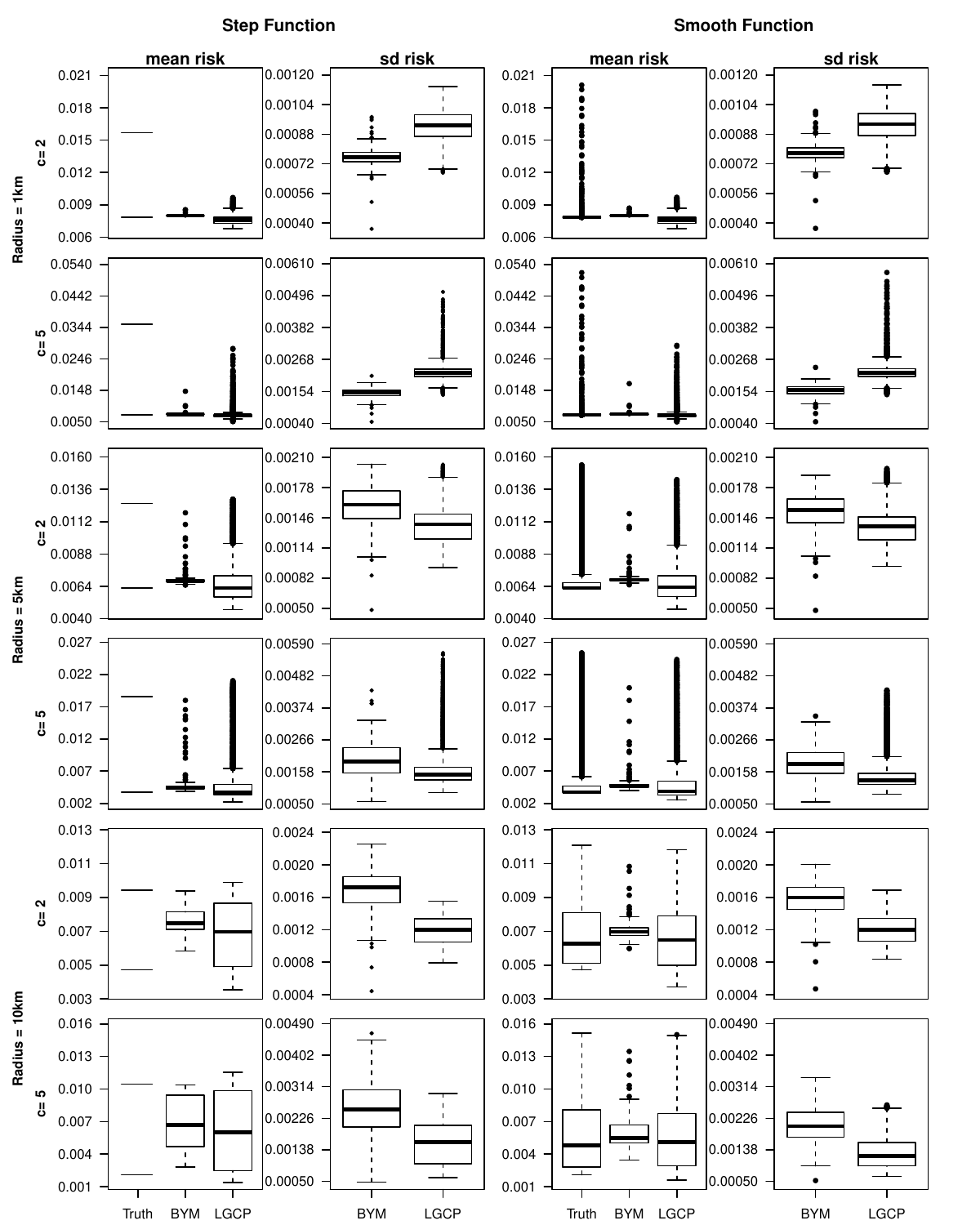}
	\caption{Spatial variation in the true risk, the mean (over the 300 simulations) of the
		posterior means and standard deviation (sd).
		The expected number of cases is kept at $5n$ ($k=5$).}\label{fig3}
\end{figure}

\begin{figure}[h!]
	\centering \includegraphics[width = 140mm]{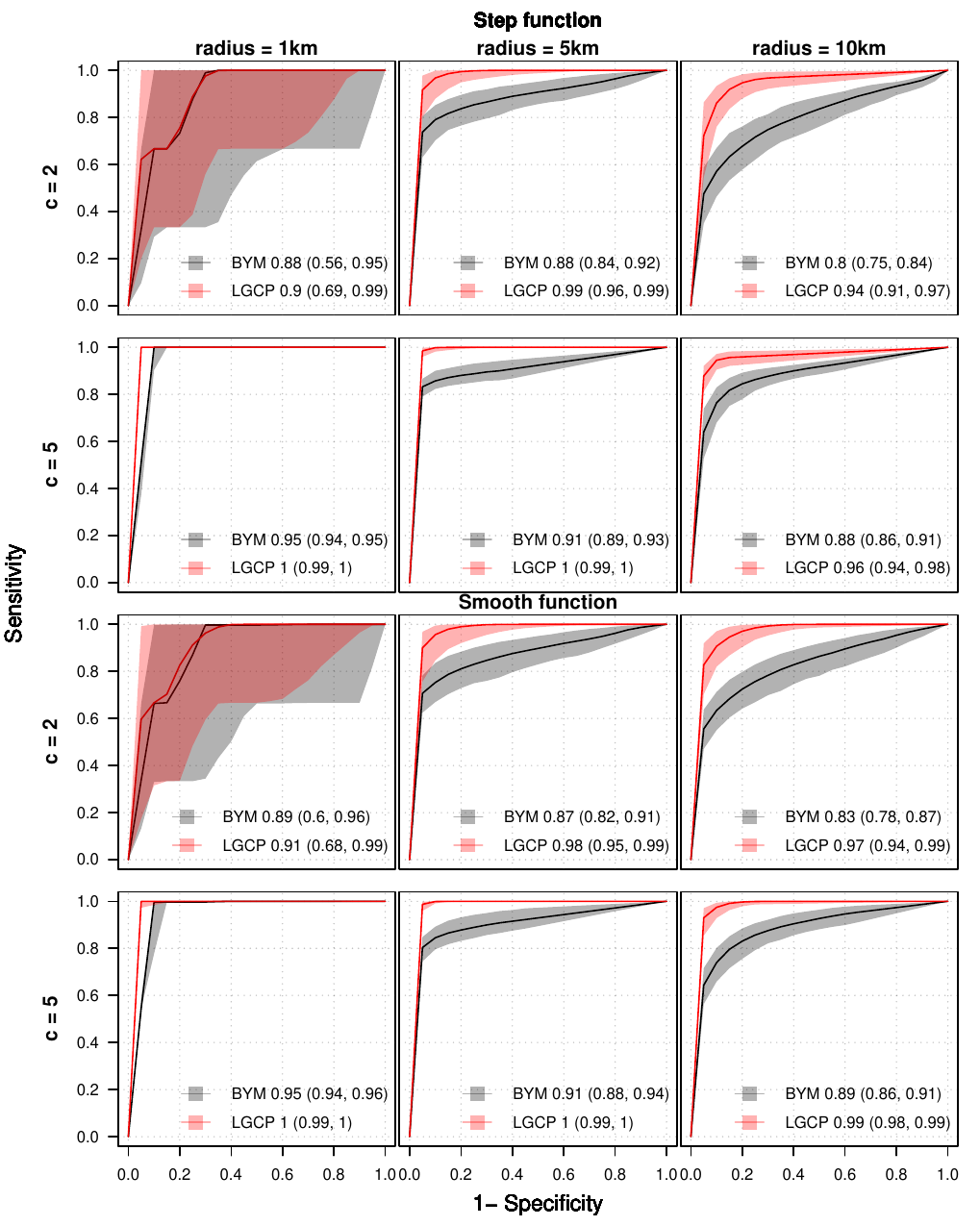}
	\caption{Pointwise median receiver operating characteristic (ROC)
		curves and their corresponding pointwise envelopes. The
		envelopes were calculated by taking the 2.5th and
		97.5th percentiles of sensitivity for given values of
		$1- $specificity across the 300 simulations. The legend shows the
		median and the 2.5th and 97.5th percentiles of the AUC across the
		simulations. The expected number of cases is $5n$ and we used area-weights.
	}\label{fig4}
\end{figure}

Table~\ref{tab2} shows the median and the $2.5$th and $97.5$th percentile
over the $300$ simulations of the area-based RMISE, evaluating the error
on the log scale ($b_g = 1$ and $\hat{R}_g = \log(\hat{\lambda}_g)$).
Regardless of the sample size or the shape of the data-generating risk
surface, LGCP outperforms BYM for large radii (\SI{10}{km}), but also for
medium radii (\SI{5}{km}) combined with high risk increases ($c=5$). In
contrast, BYM tends to outperfom LGCP in the case of small radii,
small risk increases, and when the risk surface is flat (online supplement, Table S1). The results across the scenarios are
similar when we consider the population weights or the fitted values
on the risk scale; refer to the online supplement, Tables~S2--4.

Maps of coverage probabilities are shown in Figures S6--11 in the online supplement. From these it is clear that LGCP outperforms BYM for all
data-generating scenarios with medium (\SI{5}{km}) to large (\SI{10}{km}) size of
the high risk areas. Coverage probabilities of LGCP are high both
in and outside the high-risk areas, and the only regions of poor
coverage are along the immediate boundaries of the high risk areas
in the step function scenarios. This was to be expected, given that it is
impossible for a smooth function to perfectly approximate a step
function. For the BYM, considerable extents of areas within or without
the high risk areas show sub-optimal coverage in all these scenarios.
None of the models properly capture the high risk areas when these are
confined to small circles (\SI{1}{km}). However, even for this case
the areas of low coverage are restricted to the circles for LGCP,
while they extend to the entire municipalities for BYM.

Table \ref{tab3} shows the median and $2.5$th and $97.5$th percentiles
of the coverage proportions $p_j$ (proportion of area for which the true
risks lie within the credibility regions). In line with the maps of
coverage probabilities, LGCP consistently shows a higher coverage
proportion when the data-generating process has a smooth risk surface,
while the BYM coverage proportion remains often under 95\%.
In this scenario, the only situation in
which BYM and LGCP perform similarly is when high risk areas are
small (\SI{1}{km}) and the disease rare ($k=1$). Similarly, LGCP outperforms
BYM in almost all scenarios when the underlying risk is a
step function. There are few exceptions for which BYM appears to
perform marginally better, namely for the combinations of medium or
large circles, higher risk increases, and higher incidence rates. But
as the Figures S6--11 in the online supplement show, areas of poorer
coverage for LGCPs are confined to the circular transition areas from
high to low risk. On the remaining area (both within and outside of
high risk areas) coverage probabilities tend to be high in all these
situations.

Figure \ref{fig3} shows the variation across grid cells of the mean
(over simulations) of the posterior mean and standard deviation of
estimated risk when the expected number of generated cases is set to $5n$
($k=5$). In all scenarios the geographic variability of risks
estimated by LGCP is closer to the true variability of risks compared
to estimates from BYM. This suggests a stronger tendency for shrinkage
to the mean for BYM. Thus, even when the high risk areas are small ($r=
\SI{1}{km}$), LGCP models attempt to capture these risk increases, likely
leading to greater variability in the estimates even for the areas
outside the circles. This is a plausible explanation for the poorer
performance of LGCPs in terms of RMISE for small radii and small
risk increase: The BYM model better captures the
risk outside the circles and, although it fails to capture the risks
within the circles, this yields a better RMISE because the circles
are very small. Stronger shrinkage to the mean is also a plausible
explanation for the better performance of the BYM model in the constant risk
scenario (online supplement, Figure~S12). Except in the scenarios of
small risk areas, the LGCP risk estimates tend to be more stable,
i.e.\ on average have narrower posterior distribution as shown by the
distribution of standard deviations. The results are similar for
$k=1$ and $k=10$ (online supplement, Figures~S13 and~S14).

Figure~\ref{fig4} shows the pointwise median and 95\% envelopes of
the area-based sensitivity against $1- $specificity (ROC curve). The legend
states the median and $2.5$th and $97.5$th percentiles of the
$\AUC$ over the simulations, where the expected number of generated cases
is set to $5n$. For all scenarios LGCP clearly outperforms BYM in terms
of identifying areas of high risk (AUC consistently higher). While the
two ROC curves are similar for scenarios with both small risk areas
($r=\SI{1}{km}$) and small risk increases ($c=2$), it is clearly visible that
LGCP has higher sensitivity and specificity in all other scenarios
for all the exceedance probability thresholds $q$ considered.
We observe similar results when
increasing or decreasing the number of cases or using the
population-based version of sensitivity and specificity (online supplement, Figures S15--19). For more information on the sensitivity and specificity per probability threshold $q$ refer to the online supplement, Figures S20--25.

\section{Example: Childhood leukaemia incidence in the Canton of
    Zürich}
	\label{sec:Example}

\begin{figure}[t!]
	\centering \includegraphics[width = 120mm]{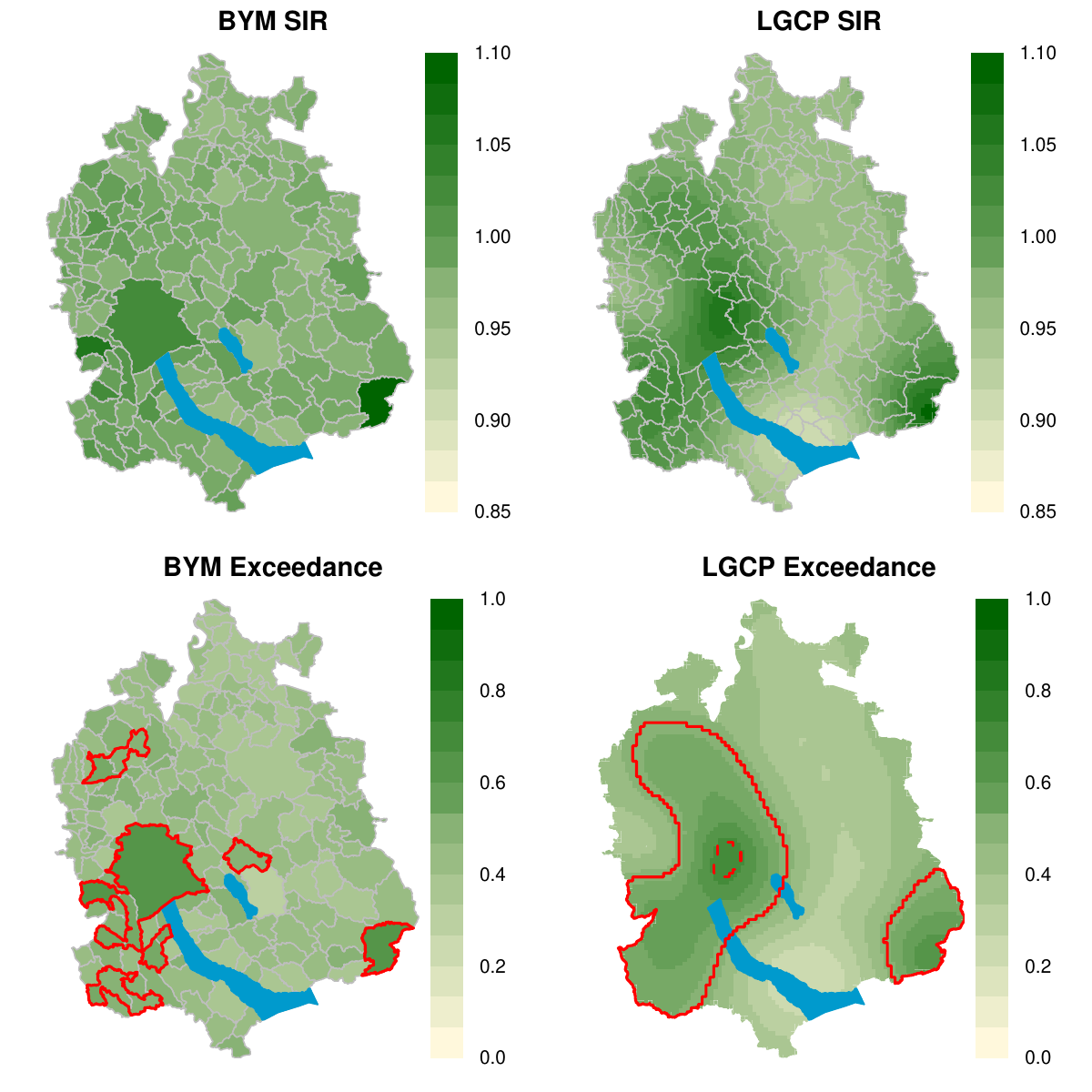}
	\caption{Posterior median standardized incidence ratio (SIR) per municipality in the BYM model (top-left panel) and per \SI{500 x 500}{m} grid cell in the LGCP model (top-right panel). The plots bellow show the
		exceedance probabilities $\Pr(\hat{\text{SIR}}>1)$, where $\hat{\text{SIR}}$ is computed
		per municipality ($\hat{\text{SIR}} = \hat{\text{SIR}}_i$) in the BYM model (bottom-left) and
		per grid cell ($\hat{\text{SIR}} = \hat{\text{SIR}}_g$) in the LGCP model
		(bottom-right panel). The red lines delimit areas where
		$\Pr(\hat{\lambda}>1)>0.5$ (solid line) and
		$\Pr(\hat{\lambda}>1)>0.75$ (dashed line).}\label{fig5}
\end{figure}

Childhood leukaemia is a rare cancer and the only established
environmental risk factor is ionising radiation in high doses
\citep{Wakeford2013}. The childhood leukaemia example is of particular
interest, as there have been a number of reports of childhood leukaemia
clusters in the literature \citep{McNally2004}. Most of these clusters
were discovered incidentally and it is not possible, in retrospect, to
judge whether they represent true deviations from a flat risk
scenario. Indeed in a recent systematic investigation of spatial
clustering in Switzerland, we found that quite remarkable aggregations
of cases are well compatible with a flat risk scenario
\citep{Konstantinoudis2017}. Disease mapping is another approach of
identifying areas of high risk, that may be more sensitive to areas of
irregular shapes and long range spatial trends.

Data for childhood leukaemia were available through the Swiss Childhood
Cancer Registry (SCCR), which is a nationwide registry with a
estimated completeness $>95\%$ since the mid 90s \citep{Schindler2015}.
For this study we used the precise geocoded locations of place at
diagnosis of the 334 registered childhood leukaemia cases diagnosed
during 1985--2015 in the canton of Zürich. Precise geocodes for all children of the general population were available through the previous decennial questionnaire-based national censuses (1990, 2000) and the annual register-based censuses beginning in 2010. The population denominator was calculated in a similar way as in \citep{Li2012a}. Briefly, we calculated the expected number of cases $E_g$ per $g$--th Voronoi cell (or municipality for the BYM) as follows:
$$E_g = \sum_{ij}\Lambda_i H_j P_{ijg}$$
\noindent where $\Lambda_i$ is the childhood leukaemia incidence in the canton of Z{\"urich} in the $i$--th year, $H_j$ are age effects corresponding to the 0--4, 5--9 and 10--15 age groups, and $P_{ijg}$ the population in the $i$--th year, $j$--th age group and $g$--th Voronoi cell (or municipality). For the non-census years we assume that the population size is the same as in the closest census year, which leads to a constant size for the years 1985--1994, 1995--2004, 2005-2010, and from 2011 an onwards we have the population available. We fitted LGCP and BYM
models using the same specifications as in the simulation study (see online supplement, Figure~S26 for prior-posterior plots of the hyperparameters). Having the expected number of cases as the denominator, adjusted for risk variations over time and age, the models estimate the standardized incidence ratio (SIR), defined as $\hat{\text{SIR}}_g = \hat{\lambda}_g/E_g$. We  mapped the SIR estimates of both models as well as the exceedance
probabilities defined as $\Pr(\hat{\text{SIR}}_g > 1)$. We highlighted areas,
for which the exceedance probabilities surpass the thresholds 0.5 and 0.75.
The sensitivity and specificity observed in our simulation study
for these thresholds are reported in Table~S5 of the online supplement.

Figure \ref{fig5} shows the fitted SIR suggested by the BYM and LGCP
models in the top panels and the exceedance probabilities in the lower
panels. Overall there appears to be little spatial variation of
childhood leukaemia SIR in the canton of Zürich. The variation of
SIR estimates from the LGCP is somewhat larger with a median SIR of 
0.98 and $[\min, \max] = [0.90, 1.10]$ compared to the variation
retrieved from the BYM model, where the median risk is 0.99 and $[\min, \max] = [0.95, 1.10]$. The map based on the BYM model is more patchy, highlighting individual municipalities that stand
out quite markedly from their neighbours. In contrast the risk surface
based on the LGCP model shows gradual changes with two spatially
coherent areas of higher risk, one near the city of Zürich and one in
the South-East of the canton. While the BYM highlights the whole
municipality of Z{\"u}rich, the LGCP shows no elevated risk in the western
part of the municipality, but locates a high risk area in the
eastern part of the municipality.  The
exceedance probability in this small area surpasses $0.75$, while the BYM
does not find any region exceeding this threshold. The estimated median of SIR increase of this particular area is $1.07$ with 95\% CI of $(0.91, 1.28)$. Assuming that there
is a real increase at this location, LGCP would have greater
sensitivity than the BYM in identifying it. This illustrates that
assuming constant risk over administrative areas may be quite
misleading. When we increased the exceedance probability to $0.80$ none of the methods reported any excess in the SIR.

We cannot know if there is true spatial variation in risk
over the period considered. The observed geographical variation in
the posterior mean of the risk is
compatible with the scenario of the simulation study, where $c=2$
and $r=\SI{1}{km}$; see online supplement, Figure S12. The observed risk increase could be spurious and an attribute to sampling variability or  imperfect spatial adjustment for person years at risk (we used population density at the census 2000, but cases were diagnosed during 1985--2015). On the other hand, the observed risk increase could be also real and an attribute to environmental factors, such as traffic related air pollutants \citep{Spycher2015},
though it is not obvious which environmental factor might be
implicated in the two areas indicated by the LGCP. Identifying
potential factors underlying the observed variation is out of the
scope of this study, and more research is required incorporating
putative risk factors.
  
\section{Discussion}
	Overall, we have found that in the framework of our study
LGCP models perform better than BYM models in
quantifying disease risk over space and in identifying areas of
high-risk. LGCP clearly outperformed BYM when risk increases and the
areas affected by these were sufficiently large to be detected. In
these situations LGCP remained superior regardless of whether the
underlying risk surface was a step function or a baseline risk plus
a Gaussian, and regardless of any changes
in the disease incidence rate. When the high-risk areas were small none of the models managed to reliably detect the increases
or quantify the risks within these areas. In these scenarios BYM
tended to produce a smaller RMISE due to a more efficient estimation
of the flat risk surface in the large remaining area. The more
reliable estimation of a flat risk surface appears to be the only
advantage of BYM over LGCP. In our example using true childhood
leukaemia incidence data from the canton of Zürich, the LGCP model
identified smooth risk increases over the continuous domain in two spatially 
coherent areas, while the map produced by BYM was patchy, with multiple
non-contiguous areas of elevated risk. Furthermore, risks estimated by
LGCP showed greater variation over space and revealed variation at the
sub-municipal level that could not be picked up by BYM.

Our results are consistent with two out of three previous studies in
the literature. Motivated by studying the lupus incidence in Toronto,
\citet{Li2012a} simulated 40 Gaussian surfaces with zero mean, keeping
the variance and roughness parameters constant ($\theta = 0.5$ and $\nu=2$) and
varying the range parameter ($\varrho = 1, 2, 3, \SI{4}{km}$). They compared the
performance of BYM and LGCP. Arguing that lupus risk is too
low, they simulated cases using stomach and lung cancer risk. They used
the mean squared error and ROC curves to examine the ability
of the models to estimate the risk and pick up areas of higher risk.
They consistently reported that the LGCP outperforms the BYM model.
\citet{Li2012} extended the LGCP model to aggregated data and compared them
with the LGCP model based on case locations and the BYM model using a similar
simulation procedure and metrics as in their previous study. They
reported that the LGCP extension on aggregated data performed better
than the BYM on aggregated data, however the LGCP on case location data was
always superior. \citet{Kang2013} simulated point data, as guided by a
previous study by \citet{Illian2012}, aggregated this data on a range of
different spatial scales and used a variety of smoothness priors to
examine the impact of spatial scale and prior in the predictive
performance of spatial models. Among the different priors were the BYM
and a Mat{\'e}rn model, which with a fine grid selection approximates an
LGCP. They conducted inference with INLA and reported mixed results
in the sense that model performance depended on the individual scenarios.

Our work has some strengths. At its heart it is an extensive simulation study using samples from a true population that yields
datasets with realistic spatial distribution of cases and persons at risk. We considered a
range of different scenarios with different sizes of high-risk areas, risk increases,
levels of urbanicity and shapes of the risk function, attempting
not to favour either of the models used for fitting. The shape of the
high-risk areas was always circular, which is an intuitive shape for
disease mapping (hot spots). This choice also provides
parsimony with respect to the parameters that need to be set and varied
(centres and radii). These strengths make our study stand out from the literature, where previous studies were based on small simulation samples (40 samples in \citep{Li2012a} in contrast with our 300) and limited scenarios (4 scenarios in \citep{Li2012a} in contrast with our 39). In addition, previous simulation studies based their scenarios on a  Mat{\'e}rn field, which is expected to favour LGCPs, \citep{Li2012} and \citep{Li2012a}. Our simulation study is based on scenarios that are unlikely to favour any of the models we selected. In addition, we selected the SPDE approach with a mesh triangulation that allows for projections on any resolution required rather than an ad-hoc grid specification. To the best of our knowledge, this is the first study that compares LGCPs with SPDE on a mesh with a BYM model. 

We need to acknowledge some limitations. Even if circles is an intuitive and parsimonious shape, more complex shapes should be considered in future studies. We also did not examine the effect of any spatially varying covariates, an issue discussed by \citet{art644}. In addition, the BYM for the current study depends on a single type of aggregation (municipalities). Presumably ZIP-code areas (the smallest areal unit in Switzerland, 268 in the canton of Zurich) would have led to preciser results. However, the choice of municipalities is justified as the smallest regional unit at which routinely collected data commonly become available while preserving data confidentiality. Our results may be sensitive to the particular setting in the canton of Z{\"u}rich (population distributions, shapes of municipalities etc.). However, we decided to focus on the canton of Z{\"u}rich for computational considerations and because it provides a representative setting with different degrees of urbanization.

The results we found are subject to the mesh specification and a
denser mesh provides more precise estimates (see
\citet{Teng2017}). 
It is tempting to assume that the failure of LGCPs to capture the risk
increases over small areas (radius \SI{1}{km}) can be attributed to the mesh
selection and the resulting loss of spatial resolution. However, this
is unlikely to be the case: We performed an ad-hoc analysis simulating
100 datasets to examine the effect of the mesh size on the RMISE, setting $b_g = 1$ and $\hat{R}_g = \log(\hat{\lambda}_g)$ and
assuming smooth risk surface and $k=5$. We selected the centroids of
\SI{500 x 500}{m} grid cells as the mesh nodes, which resulted in a mesh with $M=7563$
nodes, almost twice as many as used for the main analysis ($M=4376$;
supplementary Figure S1). The reason for choosing this regular grid is that the same grid is used for estimating the posterior risk and calculating RMISE, so that no projection of the representation~\eqref{finite} to the regular grid is required but the values $Z_i$ can be used directly. Consequently, we expect our estimates to be as close to the truth as a model of this grid size can produce. The results are reported in the online supplement, Figure S27. As expected the denser
mesh yields a more accurate risk surfaces for the LGCP
model, with the results being more pronounced for radius $=
\SI{5}{km}$. However, the denser mesh does not remove the outperformance of
BYM when radius $= \SI{1}{km}$. Increasing the mesh comes with a
considerable increase in computation time: the mean processing time of the
LGCP model in this case is approximately \SI{400}{sec} in contrast to
\SI{76}{sec} needed on average for the same scenarios under the coarser mesh
specification. This initial mesh selection was a compromise between precision and computation time across all simulations.

A more plausible explanation for the tendency of BYM to
perform better when there are just a few peaks of radius $= \SI{1}{km}$ seems to be that the large flat risk surface dominates the estimation of parameters determining variance and spatial correlation of the Gaussian field, and as a consequence these risk peaks are smoothed out.  At the same time the sensitivity estimates for both models are fairly similar (online supplement, Figures S20--25 and Table S5). These findings are in line with previous simulation studies that reported a tendency of the BYM model to oversmooth the point estimates but to perform well at overall classification of areas into higher-risk areas \citep{Best2005}.

Our results suggest that, under the given scenarios and when using
exceedance probabilities to define areas of high-risk, LGCPs may be a promising tool for cluster detection. The most popular cluster detection test is
Kulldorff's circular (or elliptic) scan \citep{Kulldorff1997, Kulldorff2006}.
However, these methods do not provide smooth risk estimates over the
domain, have difficulties in detecting clusters of irregular shapes and are
slightly conservative when there is more than one cluster in the domain.
Using a model-based approach we bypass some of these issues. However results are
expected to be sensitive to the prior specification. Furthermore the
selection of a threshold $q$ for the exceedance probabilities is
often arbitrary, creates an additional bottleneck in the analysis
and possibly multiple testing issues. For our scenarios the circular scan would be expected to perform better, as it is constructed
to be used for circular cluster detection. LGCP and other disease mapping models provide no formal test for the presence of clusters, however this avenue might be pursued in future research.  Future studies
should examine different methods for identifying high-risk areas using
LGCP or BYM models, such as excursion sets \citep{Bolin2015} or quantile
regression \citep{Padellini2018}), and compare these approaches with
Kulldorff's scan.

This study highlights the strengths of continuous domain models for
disease mapping when precise geocodes are available. However, patient confidentiality concerns are an important reason for not making such data available. Future research should seek ways to utilizing data at its maximum resolution while fully respecting
privacy concerns. In this line, it would be interesting to examine how
sensitive the results are to data perturbation (jittering) as a way
for preserving data confidentiality. Future studies should also
compare the performance of discrete and continuous domain models when
the underlying risk is linked to individual or spatial covariates. In
theory, continuous domain models should allow bypassing problems in
regression models based on discrete area units, including ecological bias
and spatial misalignment. 

A discrete approach based on administrative regions might be preferable in certain contexts. Public health policies and interventions are likely to be employed on such geographical scales and thus stakeholders and public health experts are interested in regional-based estimates. Alternatively, one could use the continuous approach and integrate the estimates on the administrative region of interest. Such integration has been previously used, but it has not been demonstrated so far to what extent this may provide preciser estimates than the discrete approach \citep{WakefieldUnder5}. In addition, the choice of the model can be driven by any information one has about the unknown spatial confounding. In aetiological studies the unknown spatial confounding is likely driven by quantities that vary continuously in space (air-pollution, temperature etc.). However in other applications, the nature of the unknown spatial confounding makes it natural to use a BYM-type specification. For instance, when a landslide occurs potential debris flow vary homogeneously within slope units, making it natural to use a BYM-type specification on the slope units \citep{lombardo2018point}. In epidemiological studies, contextual factors such as vaccine scepticism might affect individual behaviour homogeneously within a geographical region, leading again to a discrete approach \citep{Riesen2018}. Thus the main evaluation criteria for selecting methods should be based on the research question and the nature of the problem, but taking into account the benefit that can be gained by using a continuous approach.	

\section{Conclusion}

This study suggests that the use of LGCP models in
combination with point pattern data in disease mapping offers
important advantages over traditional BYM models in
combination with aggregated areal counts. LGCPs outperform BYM models in
quantifying risks and in identifying areas of high risk when the true
risk surface shows important spatial variation. In contrast BYM models
show a stronger tendency for shrinkage toward the mean and, although
being efficient in retrieving flat risk surfaces, tend to
oversmooth risk increases that occur on an intermediate spatial scale.
Our findings suggest that there are important gains to be
made from the use of continuous domain models in spatial epidemiology.

\section*{Acknowledgements}

The authors thank Dr Alex Karagiannis--Voules and Dr Haakon Bakka for
their valuable input and comments.

This work was supported by Swiss Cancer Research (4012-08-2016,
3515-08-2014). BD Spycher was supported by a Swiss National Science
Foundation fellowship (PZ00P3\textunderscore 147987).

The work of the Swiss Childhood Cancer Registry (SCCR) is supported by
the Swiss Paediatric Oncology Group (www.spog.ch), Schweizerische
Konferenz der kantonalen Gesundheitsdirektorinnen und -direktoren
(www.gdk-cds.ch), Swiss Cancer Research (www.krebsforschung.ch),
Kinderkrebshilfe Schweiz (www.kinderkrebshilfe.ch), Ernst-Göhner
Stiftung, Stiftung Domarena and National Institute of Cancer
Epidemiology and Registration (www.nicer.ch).

We thank the Swiss Federal Statistical Office for providing mortality
and census data and for the support, which made the Swiss National
Cohort (SNC) and this study possible. The work of the SNC was
supported by the Swiss National Science Foundation (grant nos.
3347CO-108806, 33CS30\textunderscore 134273 and 33CS30\textunderscore
148415).

The members of the Swiss Pediatric Oncology Group Scientific
Committee: M Ansari (Geneva), M Beck-Popovic (Lausanne), P Brazzola
(Bellinzona), J Greiner (St Gallen), M Grotzer (Zürich), H Hengartner
(St Gallen), T. Kuehne (Basel), C Kuehni (Bern), F Niggli (Zürich), J
Rössler (Bern), F Schilling (Lucerne), K Scheinemann (Aarau), N von
der Weid (Basel).

The members of the Swiss National Cohort Study Group: Matthias Egger
(Chairman of the Executive Board), Adrian Spoerri and Marcel Zwahlen
(all Bern), Milo Puhan (Chairman of the Scientific Board), Matthias
Bopp (both Zürich), Nino Künzli (Basel), Fred Paccaud (Lausanne) and
Michel Oris (Geneva).

\section*{Conflict of interest}

The authors declare that they have no conflict of interest.

\section*{Ethical standard}

Ethics approval was granted through the Ethics Committee of the Canton
of Bern to the SCCR on the 22th of July 2014 (KEK-BE: 166/2014).

\newpage

\bibliography{ref,hrue}

\end{document}